\documentclass[12pt]{article}
\usepackage{indentfirst}
\usepackage{epsfig}
\usepackage{apalike}

\begin{document}

\bibliographystyle{apalike}

\renewcommand{\baselinestretch}{1.0} \tiny\normalsize

\title{
Complexity in the Immune System
}

\author{Michael W. Deem\\
Department of Bioengineering and Department of Physics \& Astronomy\\
Rice University, Houston, TX  77005-1892\\
mwdeem@rice.edu
}

\maketitle


\begin{center}
{\bf Abstract}
\end{center}

The immune system is a real-time example of an evolving system that
navigates the essentially infinite complexity of protein sequence space.
How this system responds to disease and vaccination is discussed.
Of particular focus is the case when vaccination leads to increased
susceptibility to disease, a phenomenon termed original antigenic sin.  
A physical theory of protein evolution to explain limitations in the
immune system response to vaccination and disease is discussed, and
original antigenic sin is explained as stemming from localization of
the immune system response in antibody sequence space.  This localization 
is a result of the roughness in sequence space of the evolved antibody
affinity constant for antigen and is observed for diseases with high
year-to-year mutation rates, such as influenza.

\newpage

\section{Introduction}

Our immune system protects us against death by infection
\cite{Perelson1997}.
A major component of the immune system is generation of antibodies, protein
molecules that bind specific antigens.  To recognize
invading pathogens,
the immune system performs a search of the amino acid sequence space
of possible antibodies.  To find useful antibodies in the effectively infinite
protein sequence space, the immune system has evolved a hierarchical strategy.
The first step involves creating
the DNA sequences for B cells that code for
moderately effective antibodies through 
rearrangement of immune-system-specific
gene fragments from the genome \cite{Tonegawa,Ehrlich}.
This process is called VDJ recombination.
This combinatorial process can produce on the order of $10^{14}$ different
antibodies through recombination of pieces of antibodies.
The second step, which occurs when a specific antigen invades our body,
is somatic hypermutation.
Somatic hypermutation is the process of mutation that occurs when the
B cells that produce the antibodies divide and multiply.
Only those B cells that produce antibodies that bind the antigen with
higher affinity are propagated by this mutation and selection
process, and another name for this process is affinity
maturation.  Somatic hypermutation is essentially a search of the
amino acid sequence space at the level of individual point mutations
\cite{Griffiths,Neuberger,Tomlinson}.

Binding of an antibody to an antigen results, when the immune system
is protective, in clearance of the pathogen from which the antigen
derives.  Typical binding constants between antibodies and
antigens are on the order of $10^4$ to $10^6$  or $10^7$ l/mol \cite{Janeway},
although values as large as $10^{11}$ l/mol or greater can be produced in
experiments \cite{Wittrup}.
  Binding of antibody to antigen can prevent progression of
disease by three mechanisms.  If many antibodies bind and cover an
invading pathogen, then the pathogen is prevented from entering cells
and so fails to cause infection.  Binding of antibodies to a pathogen
may also cause phagocytic cells to ingest the pathogen.  Finally,
binding of antibodies to a pathogen may activate other components of the
immune system, such as the complement, which contains pathogen-destroying
proteins that punch holes in the membrane of the pathogen.

The space of possible antibodies is extremely large.  Considering just
the variable part of the heavy chain, there are on the order of
$20^{100} \approx 10^{130}$ possibilities.  The immune system has evolved
a hierarchical strategy for searching this vast space of possible
antibodies \cite{Janeway}.
The hierarchy is composed of two steps.  The first step
is VDJ recombination, in which the three pieces of the variable region
are joined together.  Given the many possibilities for each of these
three regions encoded within the genome of an individual, there are
on the order of $10^{12}$ possibilities of these random antibodies,
or naive sequences.  Of these, only on the order of $10^8$ are
expressed in a human individual at one point in time.
Not all of these naive sequences lead to useful antibodies.  Those
antibodies that bind proteins of the self would cause
autoimmune disease,  and the B cells that produce them are killed.  The
B cells that produce antibodies that do not bind any proteins are
left to die.  But the B cells that produce antibodies that bind antigen
multiply and divide.  As the B cells divide, the DNA that codes for the
antibodies mutates at a high rate.  The process, termed somatic
hypermutation, can be viewed as a local optimization of the initial, naive
guesses.

The consequence of an immune system response to antigen is the 
establishment of a state of memory \cite{Gray}.
Immunological memory is the ability of the immune system to respond more
rapidly
and effectively to antigens that have been encountered previously. Specific
memory is maintained in the DNA of long-lived memory B cells that can persist 
without residual antigen \cite{Black,Sprent}.

The adaptive vertebrate immune system is a wonder of modern evolution.
Under most circumstances, the dynamics of the immune system is well-matched
to the dynamics of pathogen growth during a typical infection.  Some
pathogens, however, have evolved escape mechanisms that interact in subtle
ways with the immune system dynamics.  In addition, negative interactions
between 
the immune system, which has evolved over 400 000 000 years, and vaccination,
which has been practiced for only 200 years, are possible.

 Although our immune system is highly effective, some limitations have been
reported. The phenomenon known as ``original antigenic sin'' is the
tendency for antibodies produced in response to exposure to 
influenza virus antigens to
suppress the creation of new, different antibodies in response to exposure to
different versions of the flu \cite{Fazekas1,Fazekas2}. 
Roughly speaking, the immune system responds only to the antigen
fragments, or epitopes, that are in common with the original flu virus.
As a result, individuals vaccinated against the flu
may become \emph{more} susceptible to infection 
by mutated strains of the flu than would individuals
receiving no vaccination.  

The phenomenon of original antigenic sin has since
been observed in dengue fever, human immunodeficiency virus (HIV), and
other viruses.  Dengue fever
is caused by four distinct viral strains \cite{Rigau-Perez}.
Once the immune system has memorized a response to one type of dengue,
secondary infection with a different type of dengue places patients
at the largest risk for dengue hemorrhagic
fever \cite{Halstead83,Halstead89,Vaughn}.
Thus, vaccination with one type of dengue virus can actually
\emph{increase} the rate of dengue hemorrhagic fever due
to suppression of an immune response to the three other types of
dengue virus.
Similarly,
the rapid replication of HIV, coupled with a high mutation rate, leads
to the generation
of many variants of HIV, even within
a single infected patient~\cite{Condra,Finzi,Nara}.
HIV is initially controlled by cytotoxic T lymphocytes (CTLs),
but may subsequently escape control through
mutation of the relevant T-cell epitope, since
CTLs also preferentially respond against the initial
 rather than mutated epitope \cite{Good,Klenerman}.
Original antigenic sin of the  CTL response
impedes clearance of new HIV variants
and enhances the chance that the immune system will be
unable to control the HIV infection.
The details of how original antigenic
sin works, even at a qualitative level, are unknown.

Another early example of a limitation in the immune response
came from the observation by Hoskins \emph{et al.}\
that repeated annual influenza vaccination gave worse protection
than did single vaccination among English school boys \cite{Hoskins}.
Later studies, such as that by Keitel \emph{et al.}\ observed that
repeated vaccination is more effective than
is single vaccination \cite{Keitel}. 

While the Hoskins and Keitel experiments seem inconsistent, this is
not necessarily the case.  It may be that these experiments were
performed under different conditions.  In other words, there may
be an additional order parameter that one needs to consider in
order to understand these two experiments.  This is indeed the
case.  The crucial order parameter is the number of accumulated
mutations in the influenza virus, or ``antigenic distance'' \cite{Smith}. 
As parameterized by the antigenic distance,
repeated vaccination was found to have
higher efficacy than the single vaccination
when the distance between the vaccine and the influenza
was small, whereas
repeated vaccination was found to be less effective when
the distance between the vaccine and the influenza was large.
Hoskins' and Keitel's are, thus, not necessarily inconsistent.  It is
not known, however,
why the efficacy of single and repeated vaccination should vary as
was observed in the experiments.

In this article, I offer an explanation for the reported limitations in
the immune system response using a model of protein evolution
\cite{hayoun}.
The dynamics of affinity maturation  is described
by a search in antibody sequence space
for increased binding constants between antibody and antigen.
It is shown that 
an immune system response to an antigen generates localized memory B cell
sequences.
This set of localized sequences
reduces the ability of the
immune system to respond to subsequent exposures to different
but related antigens.
It is this competitive process between memory sequences and the VDJ
recombinations of secondary
exposure that is responsible for the reported limitations in the
immune system.

\section{The Random Energy Model}
A random energy model is used to represent the interaction between
the antibodies and the influenza proteins.  This model captures
the essence of the correlated ruggedness of the interaction energy in the
variable space, the variables being the antibody amino acid sequences
and the identity of the disease proteins, and the correlations
being mainly due to the physical structure of the antibodies.  The random
energy model allows study of the sequence-level dynamics of the immune/antigen
system, which would otherwise be
an intractable problem at the atomic scale, with $10^4$ atoms
per antibody, $10^8$ antibodies per individual, $6 \times
10^9$ individuals, and many possible influenza strains.
Use of random energy theory to treat correlations in otherwise intractable
physical systems goes back at least to Bohr's random matrix theory for
nuclear cross sections \cite{Bohr1936} and has been used for
quantum chaos, disordered mesoscopic
systems, QCD, and quantum gravity \cite{Guhr1998}.
Close to the present application is the study of
spin glasses by random energy models \cite{SK1975,Derrida1980},
protein folding by coarse-grained models \cite{Wolynes1987,Gutin1993},
and evolutionary systems by NK-type models
\cite{Bogarad,Derrida1991,Weisbuch1990,Drossel2001}.

In detail, the generalized NK model used
considers three different kinds of interactions
within an antibody: interaction within a subdomain $(U^{\rm sd})$, 
interactions between
subdomains $(U^{\rm sd-sd})$, and direct binding interaction between antibody
and antigen $(U^{\rm c})$.  In the context of protein evolution,
parameters of the model have been calibrated \cite{Bogarad,Kauffman,Perelson}.  
The energy function of a protein is given by
\begin{equation}
U=\sum_{i=1}^{M} U_{\alpha_{i}}^{\rm sd} 
+\sum_{i>j=1}^{M} U_{ij}^{\rm sd-sd} 
+\sum_{i=1}^{P} U_{i}^{\rm c} \ , 
\end{equation}
where $M$ is the number of antibody secondary structural subdomains,
and $P$ is the number of antibody
amino acids contributing directly to the binding. 
The subdomain energy $U^{\rm sd}$ is 
\begin{equation}
U_{\alpha_{i}}^{\rm sd}= {1 \over \sqrt{M (N-K+1) } } \sum_{j=1}^{N-K+1}
\sigma_{\alpha_i} (a_j, a_{j+1}, \cdot \cdot \cdot, a_{j+K-1}),
\end{equation}
where $N$ is the number of amino acids in a subdomain, 
and $K$ is the range of local interaction within a subdomain.
All subdomains belong to one of $L=5$ different types
(\emph{e.g.}, helices, strands, loops, turns, and others).
The quenched Gaussian random number $\sigma_{\alpha_i}$ is different
for each value of its argument for a given subdomain type, $\alpha_i$.
All of the Gaussian $\sigma$ values have zero mean and unit variance.
The energy of interaction between secondary structures is
\begin{eqnarray}
U_{i j}^{\rm sd-sd}= \sqrt{2 \over D M (M-1)}
\sum_{k=1}^{D} && \sigma^{k}_{i j} (a^{i}_{j_1}, 
\cdot \cdot \cdot, a^{i}_{j_{K/2}};
a^{j}_{j_{K/2+1}},
\cdot \cdot \cdot, a^{j}_{j_K}) \ .
\nonumber \\ 
\end{eqnarray}
The number of interactions between secondary structures 
is set to $D=6$.
 Here $\sigma^{k}_{i j}$ and the interacting
amino acids, ${j_1, \cdot \cdot \cdot, j_K}$, are selected 
at random for each interaction $(k, i, j)$.
The chemical binding energy of each antibody amino acid to the antigen
is given by 
\begin{equation}
U_i^{\rm c}={1 \over \sqrt {P}} \sigma_i (a_i) \ .
\end{equation}
The contributing amino acid, $i$, and the 
unit-normal weight of the binding, $\sigma_i$, are chosen at random.
Using experimental results, 
$P=5$ amino acids are taken to contribute directly 
to the binding event.
Here only five chemically distinct amino acid classes 
(\emph{e.g.}, negative, positive, polar, hydrophobic, and other)
are considered
since each different type of amino acid behaves as a completely different
chemical entity within the random energy model.

The generalized NK model, while a simplified description of real proteins,
captures much of the thermodynamics of protein folding and ligand binding.
In the model, a specific B cell repertoire is represented by a specific
set of amino acid sequences.
Moreover, a specific instance of the random parameters
within the model represents a specific antigen.
An immune response that finds
a B cell that produces an antibody with high affinity constant
to a specific antigen corresponds in the model
to finding a sequence having a low energy for a specific 
parameter set.

The random character of the generalized NK model makes the
energy rugged in antibody sequence space.  The energy is,
moreover, correlated by the local antibody structure ($K = 4$),
the secondary antibody structure ($U^{\rm sd-sd}$),
and the interaction with the influenza proteins ($U^{\rm c}$).
As the immune system explores the space of possible antibodies,
localization is possible if the correlated ruggedness of the
interaction energy is sufficiently great.

\section{Parameters of the Immune System}
Since the variable region in each light and heavy chain of an
antibody is about 100 amino acids long, and most of the binding
typically occurs in one of the two chains, a 
single sequence length of 100 is chosen
\cite{Edelman,Porter}.
The value $M=10$  is chosen since there are roughly 10
secondary structures in a typical antibody and thus choose $N=10$.
The immune system contains of the order of $10^8$ B cells divided into
different specificities ($10^7$ for the mouse)
\cite{Klinman}, and 
the frequency of a specific B cell participating in
the initial immune response is roughly
1 in $10^5$ \cite{Janeway}.
Hence, $10^3$ sequences are used during an immune response.

The hierarchical strategy of the immune system is used
to search the antibody sequence space for high affinity antibodies.
Initial combination of optimized subdomains is followed by a
point mutation and selection procedure \cite{Bogarad}.
To mimic combinatorial joining of gene segments during B cell development,
a naive B cell repertoire is produced by
choosing each subdomain sequence from pools
that have $N_{\rm pool}$ amino acid segments
obtained by minimizing the appropriate $U^{\rm sd}$.
To fit the theoretical heavy-chain diversity of
$3 \times 10^{11}$ \cite{Janeway}, 
$N_{\rm pool} =  3$ sequences are chosen from among the top 300
sequences for each subdomain type.

Somatic hypermutation
occurs at the rate of roughly one mutation per variable regions of light and
heavy chains per cell division, which occurs
every 6 to 8 hours during intense cell proliferation \cite{French,Kocks}.
Hence, in the simulation, 0.5 point mutations are done per sequence, 
the best (highest affinity) $x=20$\% sequences are kept, and then 
these are amplified back up to a total of
$10^3$ copies in one round, which corresponds to $1/3$ day.
That is, the probability of picking one of the, possibly mutated, sequences 
for the next round is
\begin{equation}
p_{\rm select} = \left\{
                       \begin{array}{ll}
                          1/200,& U \le U_{200} \\
                          0,    & U > U_{200}
                       \end{array}
                 \right. \ ,
\end{equation}
where $U_{200}$ is the 200th best energy of the $10^3$ sequences after
the mutation events, and
this equation is employed $10^3$ times to select randomly the $10^3$ sequences
for the next round.
Given a specific antigen, \emph{i.e.}\ a specific
set of interaction parameters, 
 30 rounds (10 days) of point
mutation and selection are performed
in one immune response. In this way, memory B cells
for the antigen are generated. 
 
 The affinity constant is given as a function of energy,
\begin{equation}
K^{\rm eq}= \exp(a-b U) \ ,
\label{eq:affinity}
\end{equation}
where $a$ and $b$ are determined by the dynamics of the
mutation and selection process.
Affinity constants resulting from
VDJ recombination are roughly
$10^4$, affinity constants after the first response of
affinity maturation are roughly
$10^6$, and affinity constants after a second response of
affinity maturation are roughly
$10^7$ l/mol \cite{Janeway}
These values fix the selection strength, $x = 20$\%.
By comparison to the dynamics of the model, 
the values $a=-18.56$ and $b=1.67$ are obtained. 

The memory B cells, key to immunological memory, give
rapid and effective response to the same antigen due to their
increased affinity for previously-encountered antigens.
The focus is on the role of the memory cells in the immune response 
to different antigens.
The ``distance'' between a first antigen and the second antigen 
is given by the probability, $p$, of 
changing parameters of interaction within subdomain ($U^{\rm sd}$), 
subdomain-subdomain ($U^{\rm sd-sd}$),
and chemical binding ($U^{\rm c}$) terms.
Within $U^{\rm sd}$, only the subdomain type, $\alpha_i$, is changed,
not the
parameters $\sigma_\alpha$, which are probably
fixed by structural biology
and should be independent of the antigen.
This means that in the model, the primary response is always from the
same set of theoretically possible naive antibodies, analogous
to the primary response of one individual with a fixed set of 
naive possibilities from his or her genome.

\section{Immune System Dynamics}
The number of memory and naive B cells 
that participate in the immune response to the second antigen 
is estimated by
the ratio of the respective affinity constants.
From the definition of the affinity constant,
\begin{equation}
K^{\rm eq}= 
\frac{\left [ {\rm Antigen:Antibody} \right ]} {
{\left[ {\rm Antigen} \right] 
\left [ {\rm Antibody} \right ]}  } \ ,
\end{equation}
the binding probability is proportional to the affinity constant
and to the concentration of antigen-specific antibody,
which is $10^2$ times greater for the memory sequences \cite{Janeway}.
The $10^2$ times great concentration of memory antibodies is
significant, because it biases the immune response toward
the memory.
The average affinity for the second antigen
of the $10^3$ memory cells,  $K^{\rm eq}_{\rm m}$,
and that of the $10^3$ B cells from the naive repertoire
of optimized subdomain sequences,
$K^{\rm eq}_{\rm n}$, are measured.
The ratio $10^2 K^{\rm eq}_{\rm m} / K^{\rm eq}_{\rm n}$ 
gives the ratio of memory cells to naive cells.
For exposure to the second antigen, 
30 rounds (10 days) of point mutation
and selection are performed, starting with
$10^5 K^{\rm eq}_{\rm m} / (10^2 K^{\rm eq}_{\rm m}+K^{\rm eq}_{\rm n}) $ 
memory cells and
$10^3 K^{\rm eq}_{\rm n} / (10^2 K^{\rm eq}_{\rm m}+K^{\rm eq}_{\rm n})$
naive cells, since both memory and naive sequences participate
in the secondary response \cite{Berek}.

Before going in to more details of the immune system response,
I mention two possibilities for suboptimal dynamics.  The first
is localization in antibody amino-acid sequence space.
Imagine the binding energy of an antibody for this year's flu
as a function of antibody sequence.  There will be a region where
the best antibodies are located.  The binding energy of antibodies
for next year's flu will be slightly different, as the flu will
likely have mutated during the year.  The region where the best
antibodies are to be found next year will be slightly different from the
region for this year.  There may still be a local minimum for the binding 
energy of next year's flu to antibodies around
the best region for this year's flu.  If this
is the case, that means there is a barrier that separates the
region of best antibodies from one year to the next.  This,
in turn, means the
immune system must overcome this barrier in order to evolve
the antibodies during the year.  If, for example, the individual
dies from the flu before this barrier is overcome, this really is
localization of antibodies in sequence space.  Note that
the barrier may be both energetic and entropic.  That is, the mutations
required to move from one region to another may either lead to less
favorable antibodies along the way, or they may be more numerous
than would be expected simply based upon the amino acid differences
between the two optimal regions

\section{Original Antigenic Sin}

\begin{figure}
\epsfig{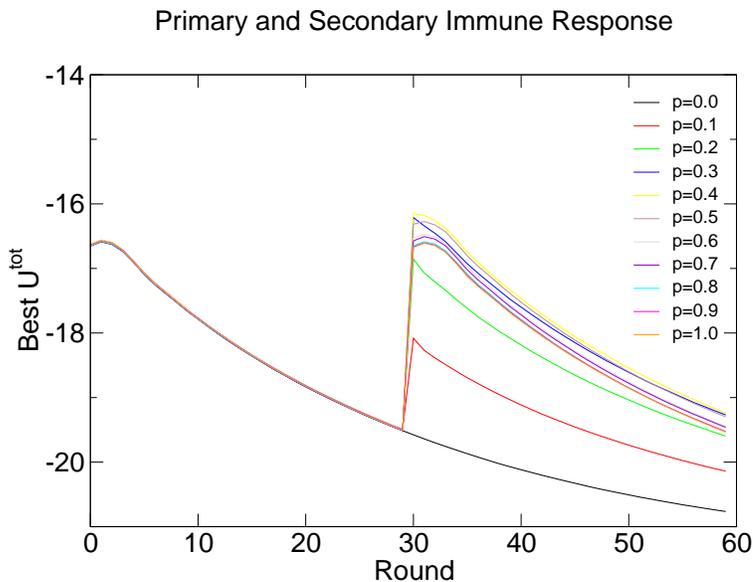}
\caption{
The dynamics of the immune response within the model.
A vaccine is given in the first year.  The individual is
exposed to the disease in the second year.  The disease 
differs from the vaccine by an amount $p$.  The results,
as with all presented here, are averaged over 5000 ensembles.
}
\label{fig:dynamics}
\end{figure}

Some specific results from the model clarify the concepts involved.
In Fig.\ \ref{fig:dynamics} is shown the dynamics of the immune response.
Again, the starting point is from $10^3$ naive sequences.  These
sequences are evolved for 30 rounds.  At each round, the sequences
are mutated at a rate of 0.5 mutations per sequence per round,
and the top 20\% are kept.  After the vaccination in year 1, 
the person is exposed to the disease in year 2.  The disease may be 
similar to or very different from the vaccine.  The degree of
similarity is measured by the parameter $p$.  The secondary response
starts with some naive sequences and some memory sequences, weighted
by the average binding constants and the relative
concentrations of each.

\begin{figure}
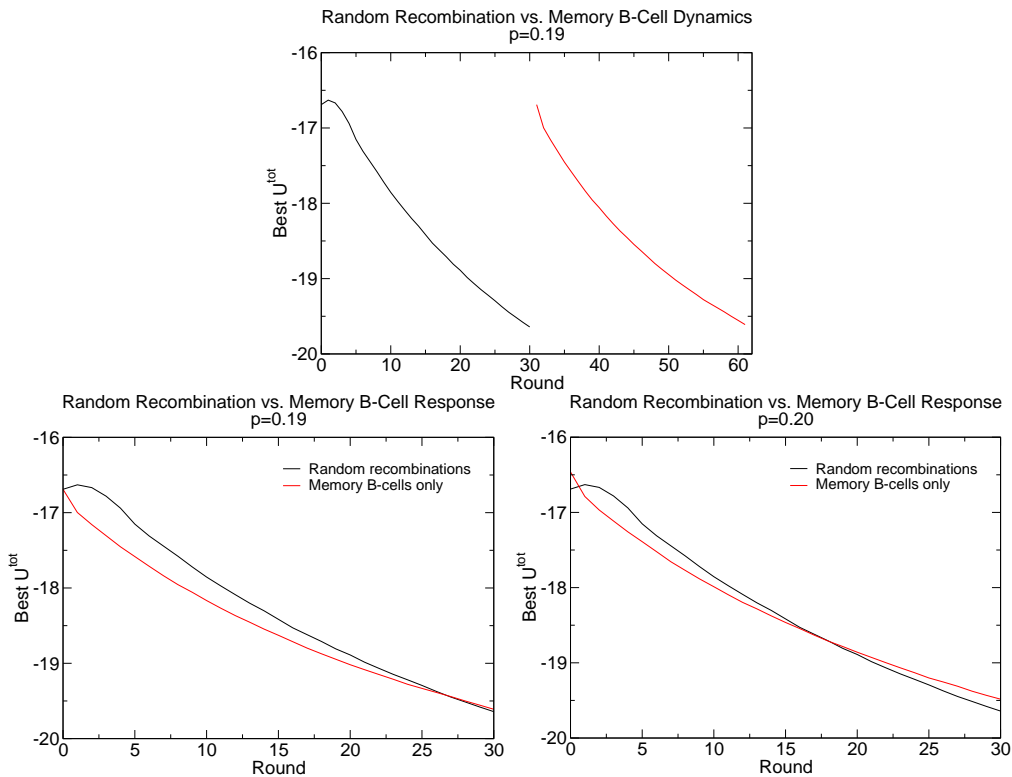

\begin{center}
\epsfig{file=memory.eps,clip=,height=2in}\\
\epsfig{file=memory2.eps,clip=,height=2in}
\epsfig{file=memory3.eps,clip=,height=2in}
\end{center}
\caption{
a) Comparison of primary and secondary response when
the secondary response is from memory sequences only.
b) Comparison when secondary response is shifted to the
   left to overlap with primary.
c) Comparison for a larger value of $p$.
}
\label{fig:local}
\end{figure}

A key point about these results is that localization does occur.
Shown in \ref{fig:local}  is the behavior of the the primary and
secondary response.  In this figure only, the secondary response
is exclusively from memory sequences.  It can be seen that in the
early part of the secondary response, the memory sequences look
superior to the naive sequences.  By the end of 30 rounds, however,
the memory sequences are worse than the naive sequences due to
an inferior ability to evolve.

\begin{figure}
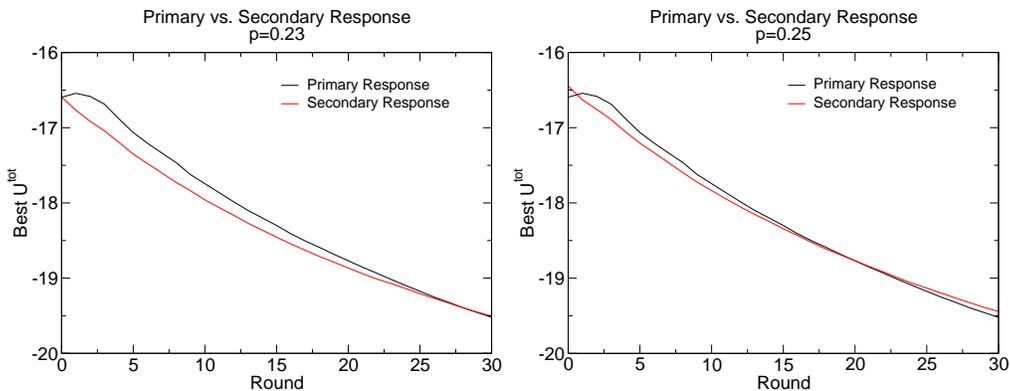

\begin{center}
\epsfig{file=dynamics1a.eps,clip=,height=2in}
\epsfig{file=dynamics1b.eps,clip=,height=2in}
\end{center}
\caption{
a) Comparison of primary and secondary response when
  the secondary response is from memory and naive sequences.
b) Comparison for a larger value of $p$.
}
\label{fig:local2}
\end{figure}

The results from \ref{fig:local} are somewhat academic, in that the
true secondary response was not used.  In figure
\ref{fig:local2}  is shown the  response when the true secondary
response is used.  It is again observed that the memory sequences
look  promising at the beginning of the response, but they are
unable to evolve as well as naive sequences.  In other words,
the presence of the vaccine that generated the memory response
leads to a worse response than if there were no vaccine.  This is
a demonstration of original antigenic sin.

\begin{figure}
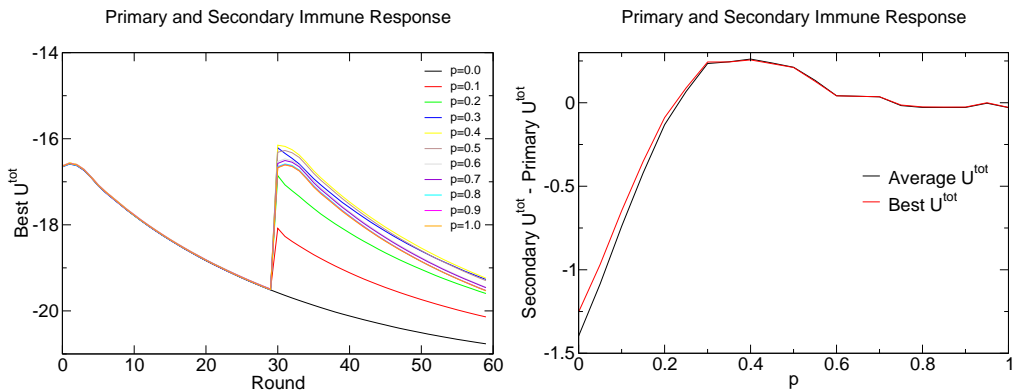

\begin{center}
\epsfig{file=dynamics2.eps,clip=,height=2in}
\epsfig{file=primarysecondary2.eps,clip=,height=2in}
\end{center}
\caption{
a) The model dynamics.  The effectiveness of vaccination is
   determined by comparing the secondary response at round 60 with the
   primary response at round 30.
b) The energy evolved in the secondary response minus that evolved in the
   primary response.
}
\label{fig:local3}
\end{figure}

It is of interest to measure the degree of original antigenic
sin as a function of the mutation of the flu.  That is,
one wants to compare the secondary response at day 10 with a
primary response at day 10.  Shown in 
Fig.\ \ref{fig:local3} is this comparison.  The secondary
response is better than the primary when the vaccine and
flu are similar.  When the vaccine and flu are very
different, the secondary response uses mostly naive sequences,
and it becomes like a primary response.   In between, there is the
region of original antigenic sin.

\begin{figure}
\begin{center}
\epsfig{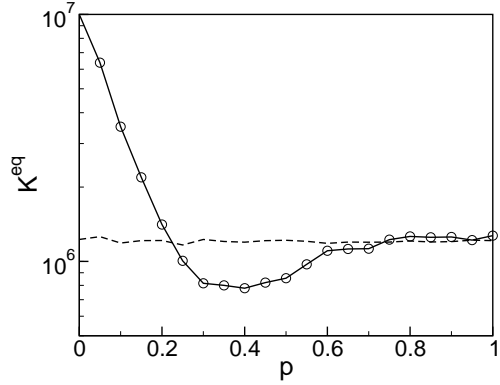}
\end{center}
\vspace{0.5cm}
\caption{
The evolved affinity constant to a second antigen after exposure to
an original antigen that differs by probability $p$ (solid line).
The dotted line represents the affinity constant without previous exposure.
The affinity constant is generated by exponentiating,
as in Eq.\ (\ref{eq:affinity}),
the average of the best binding energy.
}
\label{fig:localization}
\end{figure}

It is of interest to measure the degree of original antigenic
sin as a function of the mutation of the flu.  That is,
Fig.\ \ref{fig:localization} shows the evolved affinity constant
to a second antigen if the exposure to a first antigen 
exists (solid line) or not (dashed line) as a function of the difference
between the first and second antigen, $p$, or
``antigenic distance'' \cite{Smith}.
When the difference is small, the exposure to a first antigen
leads to higher affinity constant than without exposure, which is
why immune system memory and
vaccination is effective.
For a large difference, the antigen encountered in the first exposure
is uncorrelated with that in the second exposure, and so immune system 
memory does not play a role.
Interestingly, the immunological memory from the first exposure
actually gives worse protection, \emph{i.e.}\  a
lower affinity constant, for intermediate differences---which is
original antigenic sin.

\begin{figure}
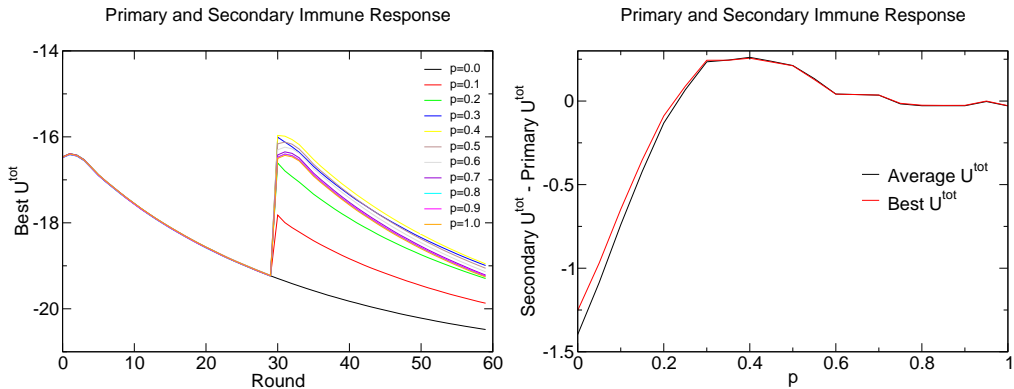

\begin{center}
\epsfig{file=dynamics3.eps,clip=,height=2in}
\epsfig{file=primarysecondary3.eps,clip=,height=2in}
\end{center}
\vspace{0.5cm}
\caption{
a) The model dynamics with a repertoire size of 500 instead of 1000.
   The effectiveness of vaccination is
   determined by comparing the secondary response at round 60 with the
   primary response at round 30.
b) The energy evolved in the secondary response minus that evolved in the
   primary response.
}
\label{fig:localization2}
\end{figure}

The dynamics of the immune response (Fig.\ \ref{fig:local3})
depend upon the constants of Nature, \emph{i.e.}\
the parameters of the model.
For example, in an organism with a smaller immune system, such as the mouse,
there are fewer starting sequences, and
less favorable binding constants are measured in the same
number of rounds: A factor of 0.5 reduction
in the number of starting sequences leads to a 0.64 reduction in the evolved
binding constant, but a similar degree of original antigenic sin as in
Fig.\ \ref{fig:local3}, as shown in
Fig.\ \ref{fig:localization2}.  The original antigenic sin curve 
looks very similar for the two repertoire sizes.  The only significant
difference is that the absolute values of the evolved energies are
slightly worse for the smaller repertoire size, due to the reduction
in diversity.

\begin{figure}
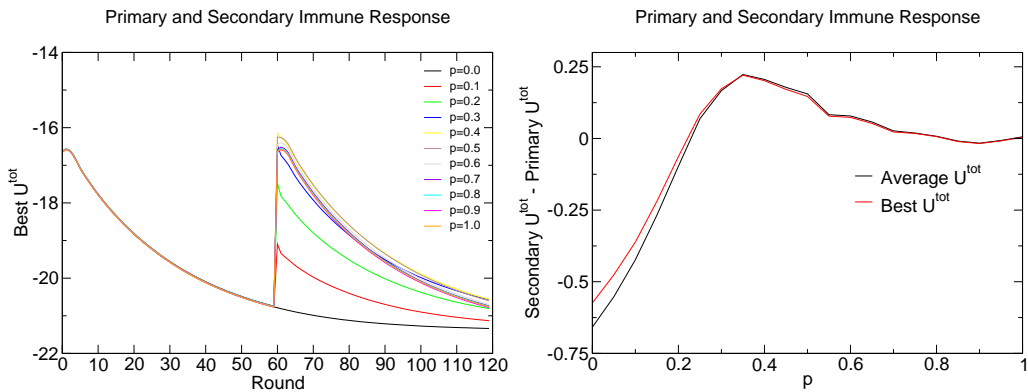

\begin{center}
\epsfig{file=dynamics4.eps,clip=,height=2in}
\epsfig{file=primarysecondary4.eps,clip=,height=2in}
\end{center}
\vspace{0.5cm}
\caption{
a) The model dynamics with 60 rounds instead of 30.
   The effectiveness of vaccination is
   determined by comparing the secondary response at round 120 with the
   primary response at round 60.
b) The energy evolved in the secondary response minus that evolved in the
   primary response.
}
\label{fig:localization3}
\end{figure}

On the other hand, if more
rounds are performed, better binding constants are found in the primary and
secondary responses, but the secondary response is not as improved over
the primary response as when
using 30 rounds. This is because the evolved sequences are becoming
more localized in ever deeper wells.  The degree of original antigenic sin is,
however, similar in the range of 30 to 60 rounds per response.  
The model dynamics are shown in Fig.\ \ref{fig:localization3}.
While 60 rounds is really too many for a single response, this experiment
could represent administration of a vaccine twice in year one, 
administration of a vaccine once in year two, and exposure to
the disease in year two.  In this case, one imagines
that the two vaccines in year one are
identical and that the vaccine and disease in year two
are identical.

Similarly, if the roughness of the
energy upon disease mutation is increased,
for example by assuming that mutation of the influenza actually changes
the $\sigma_\alpha$, the degree of original antigenic sin increases
substantially, by a factor of 2, because the barriers between the
regions of localization in sequence space are increased.
On the other hand, if the 
concentration of the memory cells is decreased, the contribution of the
memory cells to the dynamics is reduced, and the original antigenic
sin phenomenon decreases, almost disappearing when the memory and
naive antibody concentrations are equal.
Larger selection strengths ($x < 20$\%) cause
more localization and original antigenic sin, in shallower wells for
small $x$, and smaller strengths ($x > 20$\%) lead to less 
evolution.

\begin{figure}
\begin{center}
\epsfig{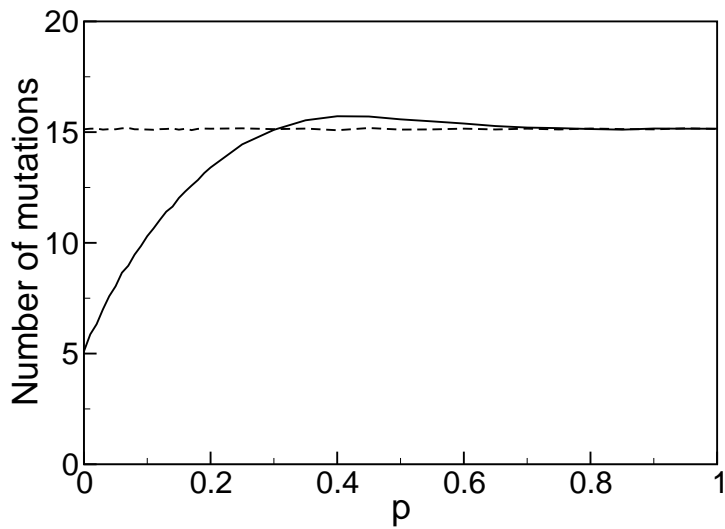}
\end{center}
\vspace{0.5cm}
\caption{
The number of mutations that occur in the best evolved antibody during the
primary (dashed) and secondary (solid) responses.
}
\label{fig:nmuts}
\end{figure}

The average number of mutations leading to the best antibody,
a measure of the localization length when original antigenic sin
occurs, is 15 for the first response and rises from 5 to 15
for the second response in the range $0 \le p < 0.30$,
as shown in Fig.\ \ref{fig:nmuts}.
Interestingly, the average number of mutations rises slightly above 15
in the range $0.30 \le p < 0.70$, indicating that in the original
antigenic sin region
more mutations are necessary for the compressed ensemble
of memory sequences from the primary response to evolve to a suitable
new state in the secondary response.  This result implies that there
is at least some entropic character to the
 barrier between the optimal regions of antibody sequence space from
one year to the next.

\begin{figure}
\begin{center}
\epsfig{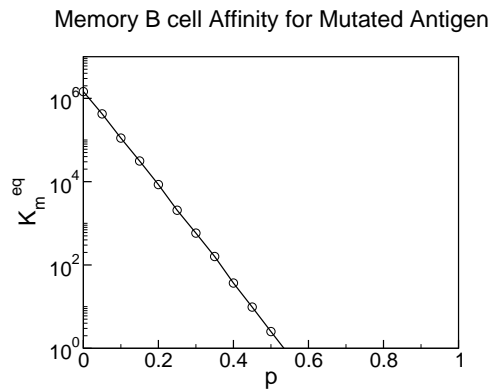}
\end{center}
\vspace{0.5cm}
\caption{
The affinity constant of an antibody as the antigen to which it
evolved against is mutated.
}
\label{fig:cross}
\end{figure}

For small values of $p$, $p < 0.19$, the memory B cells produce antibodies
with higher affinities, $K^{\rm eq}_{\rm m} > 10^4$ l/mol,
 for the new antigen than do naive B cells.
 The binding constant of the memory antibodies steadily
decreases with $p$, reaching the non-specific value of 
$K^{\rm eq}_{\rm m} = 10^2$ at $p = 0.36$, as shown in
Fig.\ \ref{fig:cross}.  
Interestingly,
this is less than the range to which 
original antigenic sin extends,
$0.23 < p < 0.60$.
 These model predictions are in good
agreement with experimental data
on cross-reactivity, which ceases to occur
when the amino acid sequences are more than 33--42\% different
\cite{East}.
This comparison also shows that the parameter $p$, which in the
model is the probability with which the parameters in the random
energy model are changed, can be interpreted
as the degree of amino acid difference.

\begin{figure}
\begin{center}
\epsfig{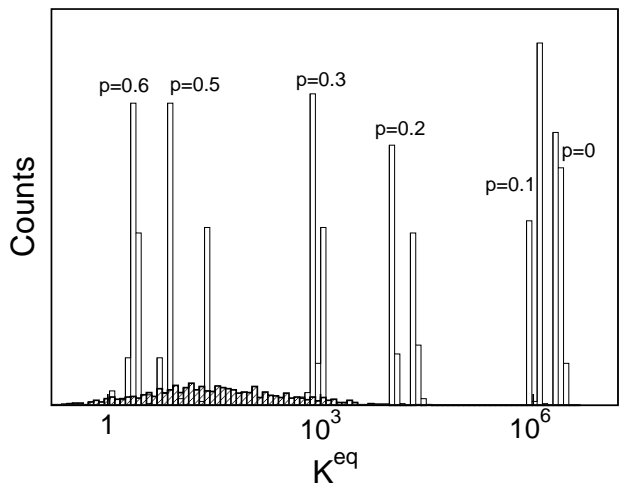}
\end{center}
\vspace{0.5cm}
\caption{
The affinity distribution of antibodies from
naive B cells (hatched) or memory cells
(open) for the antigen of second exposure.
}
\label{fig:histogram}
\end{figure}

\begin{figure}
\begin{center}
\epsfig{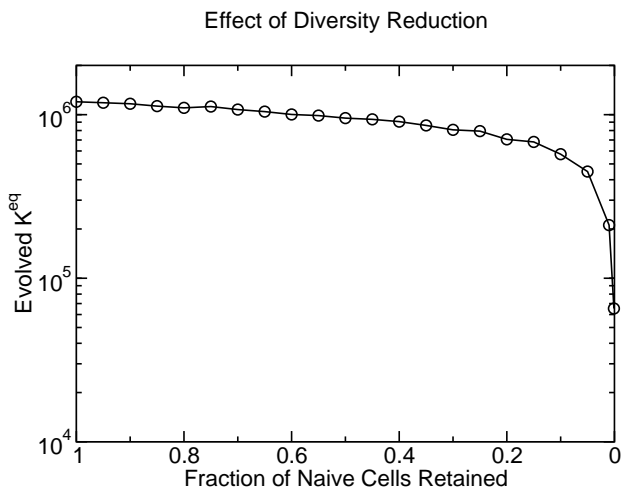}
\end{center}
\vspace{0.5cm}
\caption{
The evolved binding constant when only the top fraction
of naive sequences are kept in the dynamics.
}
\label{fig:kept}
\end{figure}

The ineffectiveness of immune system memory over a window of $p$ values
can be understood from the localization of memory B cell sequences.
Figure \ref{fig:histogram} displays distributions 
of memory and naive affinity constants
for the second antigen.
Notice that the memory sequences are highly homogeneous and lack diversity
compared to the naive sequences.  Indeed, original antigenic sin
arises mainly because the memory sequences from the primary response
suppress use of naive sequences in the tail of the distribution for the
secondary response.  Although those naive sequences initially look
unpromising, they may actually evolve to sequences
with superior binding constants.  
Figure\ \ref{fig:kept} 
illustrates this phenomenon.
Interestingly, when half of the
distribution is removed, the reduction in the binding constant is just about
that which occurs in original antigenic sin, Fig.\  \ref{fig:localization}.

The phenomenon of original antigenic sin, then, arises because the
memory sequences cut off the tail of the naive sequences with poor
binding constants.  In most cases, \emph{i.e.}\ for most values of
$p$, this is an appropriate strategy.  In some cases, however, those
unpromising looking sequences actually lead to a better evolved
binding constant than do the memory sequences.  It is these cases
in which original antigenic sin occurs.  This mechanism can be
tested by allowing evolution of only the top fraction of naive
sequences.  That is, the top fraction of the sequences is kept,
overwriting the bottom fraction with the worst kept sequence.
The repertoire size is still $10^3$, but the diversity is
the fraction kept times $10^3$.  Shown in Fig.\ \ref{fig:kept}
is the result.  It is seen that if only the best sequences is kept,
a very poor evolved antibody results.  If only the 50\% best
sequences are kept, the best evolved antibody is 
worse by just about the depth of the
original antigenic sin valley in Fig.\ \ref{fig:localization}.

\section{Single and Repeated Vaccination}

\begin{figure}
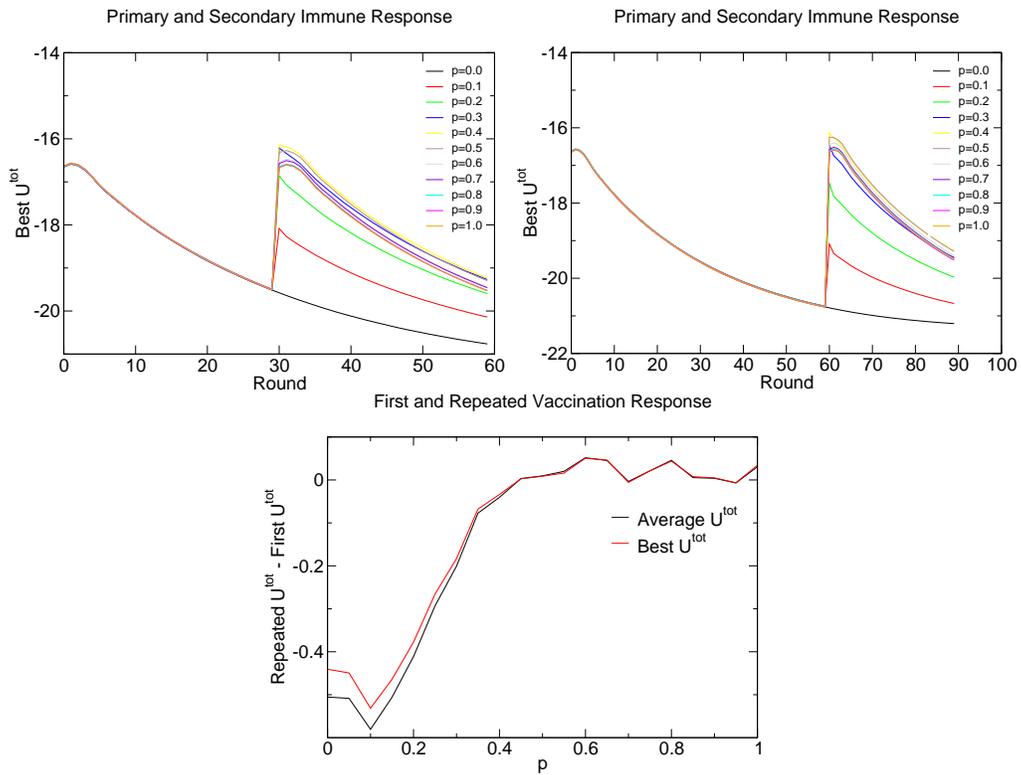

\begin{center}
\epsfig{file=dynamics2.eps,clip=,height=2in}
\epsfig{file=dynamics5.eps,clip=,height=2in}
\epsfig{file=repeated.eps,clip=,height=2in}
\end{center}
\vspace{0.5cm}
\caption{
a) Dynamics with single vaccination.
b) Dynamics with repeated vaccination.
c) Comparison between single and repeated vaccination.
}
\label{fig:ratio}
\end{figure}

\begin{figure}
\begin{center}
\epsfig{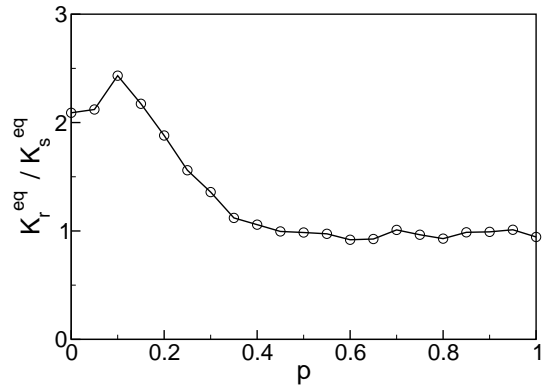}
\end{center}
\vspace{0.5cm}
\caption{
The ratio of the affinity constants to influenza
after repeated vaccination and single vaccination.
}
\label{fig:ratio2}
\end{figure}

The model is also used to address the effectiveness of
repeated annual influenza vaccination.
Repeated vaccination will lead to a more effective response against
closely related antigens.  The
ratio of the affinity constant to influenza after repeated
vaccination, $K^{\rm eq}_{\rm r}$,
to that after single vaccination, $K^{\rm eq}_{\rm s}$, 
should be greater than unity for small $p$ and converge to
unity for $p \to 1$.
Repeated vaccination generates highly localized memory cells
that have higher affinity constants to both the vaccine and the
influenza if $p$ is small.
Results for our model are shown in Figs.\ \ref{fig:ratio}
and \ref{fig:ratio2}.
The decay of the ratio $K^{\rm eq}_{\rm r}/K^{\rm eq}_{\rm s}$ 
to unity is in accord with experimental data
\cite{Smith}.  Interestingly, for our present vaccination protocol,
no localization is observed, which is consistent with most of the
modern experiments and with current public health policy that 
recommends for repeated, annual vaccinations of individuals at risk.

\section{Summary}
In summary, the generalized NK model is shown to successfully
model immune system
dynamics.  A localization mechanism for the original antigenic sin phenomenon 
observed in the flu is explained.
Localization of antibodies in the amino acid sequence space around
memory B cell sequences is shown to lead to a decreased ability of the
immune system to respond to diseases with year-to-year mutation rates
within a critical window.
This localization occurs because of the ruggedness of the evolved
affinity constant in amino acid sequence space.

From the model dynamics, it is found that
memory sequences can both out compete the non-vaccinated
immune response and become trapped in local minima.
Memory sequences with affinity constants 
initially superior to those from 
naive sequences can be selected by the dynamics,
and these memory sequences can lead to poorer evolved affinity constants, 
to the detriment of the immune system for intermediate disease
mutation rates. 
 Interestingly, due to the high diversity of the
naive sequences, the localization phenomenon extends even beyond the
region of cross-reactivity between memory antibodies and mutated 
antigen.  

These results suggest several implications for vaccination strategy:
the difference between vaccinations administered on a repeated basis should
be as great as practicable, and suppression of the memory B cell response 
may be helpful during vaccination against highly variable antigens.

\section*{Acknowledgements}
The author thanks Jeong-Man Park for stimulating discussions.
This research was supported by the National Science and 
Camille \& Henry Dreyfus Foundations.

\bibliography{immune_cce}

\end{document}